\documentclass[twoside,fleqn]{article}

\usepackage{espcrc2}
\input epsf
\newcommand{\psfile}[1]{
  \setlength{\epsfxsize}{\columnwidth}
  \epsfbox{#1}\vspace{-15pt}
}

\newcommand{\beq}{\begin{equation}}
\newcommand{\eeq}{\end{equation}}
\newcommand{\beqn}{\begin{eqnarray}}
\newcommand{\eeqn}{\end{eqnarray}}
\newcommand{\beqns}{\begin{eqnarray*}}
\newcommand{\eeqns}{\end{eqnarray*}}
\newcommand{\vs}{\\[0.3cm]\indent}
\newcommand{\vsp}{\vspace{0.25cm}}
\newcommand{\vsm}{\vspace{-0.3cm}}
\newcommand{\hm}{\hspace{-0.05cm}}
\newcommand{\hsm}{\hspace{-0.2cm}}
\newcommand{\intl}{\int\limits}
\newcommand{\ointl}{\oint\limits}
\newcommand{\mc}{\multicolumn}
\def\NP{{\it Nucl. Phys.}}
\def\PL{{\it Phys. Lett.}}
\def\PR{{\it Phys. Rev.}}
\def\PRep{{\it Phys. Rep.}}
\def\PRL{{\it Phys. Rev. Lett.}}
\def\NIM{{\it Nucl. Inst. Meth.}}
\def\ZP{{\it Z. Phys.}}
\def\TAU{Talk given at the TAU96 Conference, Colorado, 1996}

\def\QCD{Talk given at the QCD96 Conference, Montpellier, 1996}
\def\CPC{{\it Comp. Phys. Comm.}}
\def\ea{{\it et al.}}
\def\Cl{Collaboration}
\def\tauto{$\tau^{-\!}\rightarrow\,$}
\def\nut{$\,\nu_\tau$}
\def\piz{$ \pi^0 $}
\def\pipiz{$ \pi^-\pi^0 $}
\def\pidpiz{$ \pi^-2\pi^0 $}
\def\tpidpiz{$ 2\pi^-\pi^+2\pi^0 $}
\def\pitpiz{$ \pi^-3\pi^0 $}
\def\tpi{$ 2\pi^-\pi^+ $}
\def\tpipiz{$ 2\pi^-\pi^+\pi^0 $}
\def\pitpiz{$ \pi^-3\pi^0 $}
\def\tpitpiz{$ 2\pi^-\pi^+3\pi^0 $}
\def\piqpiz{$ \pi^-4\pi^0 $}
\def\pifpiz{$ \pi^-5\pi^0 $}
\def\fpi{$ 3\pi^-2\pi^+ $}
\def\fpipiz{$ 3\pi^-2\pi^+\pi^0 $}

\def\htpiz{$ {\mathrm h}^-3\pi^0 $}

\def\GeVM{~GeV$/c^2$}
\def\GeVM2{~GeV$^2/c^4$}

\def\MeVM{~MeV$/c^2$}

\def\MeVE{~MeV}
\def\pc{$\%$}

\def\sf{spectral function}
\def\sfs{spectral functions}

\def\Sfs{Spectral functions}

\def\SFS{SPECTRAL FUNCTIONS}
\def\as{$\alpha_s$}
\def\asm{$\alpha_s(M_\tau)$}
\def\ee{$e^+e^-$}

\def\rs{\raisebox{1.5ex}[-1.5ex]}
\def\pms{$\,\pm\,$}
\def\TT{$\times10^{-3}$}
\def\ie{{\it i.e.}} 
\def\eg{{\it e.g.}} 
\def\via{via} 
\title{Vector and Axial-Vector Spectral Functions and QCD}
 
\author{Andreas H\"ocker$^{\;\mathrm a,}
$ 
  \address{$^{\mathrm a}$Laboratoire de l'Acc\'el\'erateur Lin\'eaire, \\
           IN2P3-CNRS et Universit\'e de Paris-Sud, F-91405 Orsay, France }
}
 
\begin{document}

\begin{abstract}
We present new results for the $\tau$ hadronic \sfs\ analysis
using data accumulated by the ALEPH detector at LEP during the years 
1991--94. In addition to the vector \sf, the axial-vector \sf\ and, 
separately, the \tauto\tpi\nut\ as well as the \tauto\pidpiz\nut\ \sfs\ 
are determined from their respective unfolded, \ie, physical 
invariant mass spectra. The \sfs\ are applied to QCD chiral sum rules 
in order to extract information about saturation at the $\tau$ mass 
scale. Using the the semi-leptonic $\tau$ decay rate for vector and 
axial-vector currents in addition to spectral moments, we obtain precise 
measurements of the strong coupling constant \asm\ and the 
contributing non-perturbative power terms. The evolution to the 
Z mass yields \as$(M_{\mathrm Z})$\,=\,0.1219\pms0.0019.
\end{abstract}
 

\maketitle

%
%
\section{INTRODUCTION}

\Sfs\ of hadronic $\tau$ decays are the key objects for various interesting
studies concerning resonance structure analysis, the {\it Conserved Vector 
Current} hypothesis (CVC) and QCD tests involving strong sum rules and the 
measurement of \as. The ALEPH measurement of the vector 
\sfs\ and related topics was presented at the same conference by 
R.~Alemany~\cite{ricard} (see also~\cite{aleph_vsf}).
\vs
In this article we deal with new data of the \sfs\ from non-strange,
axial-vector hadronic $\tau$ decays measured by the ALEPH Collaboration. 
We create the sum and the dif\/ference of vector and axial-vector \sfs\ to 
get access to information about the saturation of QCD sum rules at the $\tau$ 
mass scale. As an application, we determine the electric polarisability
of the pion from chiral QCD constraints. In calculating the semi-leptonic 
$\tau$ widths and so-called spectral moments of the vector and axial-vector 
currents, we independently f\/it the strong coupling constant 
\asm\ and the respective non-perturbative contributions of the 
{\it Operator Product Expansion} (OPE) power series~\cite{svz,bnp} using 
the framework already exploited
in previous analyses~\cite{aleph_as,laurent,cleo_as}. We f\/inally perform 
a combined f\/it of all components in order to obtain the best value of
\asm.

%
%
\section{\SFS\ OF VECTOR AND AXIAL-VECTOR CURRENTS}

The measurement of the non-strange $\tau$ vector (axial-vector) 
current \sfs\ requires the selection and identif\/ication of $\tau$ 
decay modes with a G-parity G=$+$1 (G=$-$1), \ie, hadronic channels 
with an even (odd) number of neutral {\it or} charged pions. The isovector 
\sf\ $v_{1,\,{V^-}}$ ($a_{1,\,{A^-}}$) of a vector (axial-vector) $\tau$ 
decay channel ${V^-}$\nut\ (${A^-}$\nut)
is obtained by dividing the normalized invariant mass-squared distribution 
$(1/N_{V/A^-})(d N_{V/A^-}/d s)$ for a given hadronic mass $\sqrt{s}$ by the 
appropriate kinematic factor (throughout this article, charge conjugate 
states are implied):
\beqn\label{eq_sf}
   \lefteqn{v_{1,\,{V^-}}/a_{1,\,{A^-}}(s) \;=\;} \nonumber \\
     & &   \frac{M_\tau^2}{6\,|V_{ud}|^2\,S_{\mathrm{EW}}}\,\,
              \frac{B(\tau^-\rightarrow {V/A^-}\,\nu_\tau)}
                   {B(\tau^-\rightarrow e^-\,\nu_\tau\bar{\nu}_e)} \nonumber \\
     & &   \times\frac{d N_{V/A^-}}{N_{V/A^-}\,ds}\,
              \left[ \left(1-\frac{s}{M_\tau^2}\right)^{\!\!2}\,
                     \left(1+\frac{2s}{M_\tau^2}\right)
              \right]^{-1}\hspace{-0.3cm},
\eeqn
where $|V_{ud}|\,=\,0.9752\,\pm\,0.0007$ denotes the CKM weak mixing matrix 
element~\cite{pdg} and $S_{\mathrm{EW}}\,=\,1.0194$ accounts for electroweak 
second order corrections~\cite{s_ew}. The $\tau$ mass 
$M_\tau\,=\,1776.96^{+0.31}_{-0.27}$\MeVM\ is taken from the recent BES 
measurement~\cite{bes}. The \sfs\ are normalized by the ratio of the  
respective vector/axial-vector branching fraction
$B(\tau^-\rightarrow {V^-/A^-}\,\nu_\tau)$ to the branching fraction of the
electron channel 
$B(\tau^-\rightarrow e^-\bar{\nu}_e\nu_\tau)$\,=\,17.79\pms0.04~\cite{br_e},
where the latter is additionally constrained \via\ universality from
$B(\tau^-\rightarrow \mu^-\bar{\nu}_\mu\nu_\tau)$ and the $\tau$ lifetime.
Note that our def\/inition of the $\tau$ \sfs\ dif\/fers from the one 
used in~\cite{donoghue} by an additional factor $4\pi^2$.
\vs
Assuming unitarity (which implies the {\it optical theorem}) and analyticity 
to hold, the \sfs\ of hadronic $\tau$ decays are related \via\
dispersion relations to the imaginary parts of the two-point 
correlation functions $\Pi_{ij,U}^{\mu\nu}(q)\,=\,i\int d^4x\,e^{iqx}
\langle 0|T(U_{ij}^\mu(x)U_{ij}^\nu(0)^\dag)|0\rangle$ of vector 
($U\equiv V\,=\,\bar{\psi}_j\gamma^\mu\psi_i$) or axial-vector ($U\equiv 
A\,=\,\bar{\psi}_j\gamma^\mu\gamma_5\psi_i$) colour-singlet quark currents 
in corresponding quantum states (see, \eg,~\cite{bnp,pichtau94}).

%
%
\section{THE MEASUREMENT PROCEDURE}

The measurement of the \sfs\ def\/ined in Eq.~(\ref{eq_sf})
requires the determination of the physical invariant mass-squared 
distribution. The details of the analysis are reported
in~\cite{aleph_vsf}. A description of the ALEPH detector and its 
performance is published in~\cite{aleph_det}. 

In the following, we will brief\/ly outline the important steps of the 
measurement procedure:
\begin{table*}[thb]
\setlength{\tabcolsep}{1.0pc}
\begin{center}
\parbox{13cm}
{
 \caption[.]{\label{tab_va}
                Vector and axial-vector hadronic $\tau$ decay modes with 
                their contributing branching fractions. According to~\cite{michel}, 
                the K$\bar\mathrm K\pi$ channels are assumed to be 
                (78$^{+22}_{-28}$)\pc\ vector and (25\pms25)\pc\ axial-vector, 
                while the errors are 100\pc\ anti-correlated.  
                The K$\bar{\mathrm K}\pi\pi$\nut\ modes are conservatively 
                assumed to have (50\pms50)\pc\ vector and axial-vector parts.}
\vspace{0.3cm}
}
\begin{tabular}{|c|c||c|c|} \hline
\mc{1}{|c}{Vector}     & \mc{1}{c||}{BR (in \pc)}   
                                       & \mc{1}{c}{Axial-Vector}
                                                       & \mc{1}{c|}{BR (in \pc)} \\
   \hline\hline
\pipiz\nut\            & 25.35\pms0.19 & $\pi^-$                & 11.23\pms0.16  \\
\pitpiz\nut\           &  1.17\pms0.14 & \pidpiz\nut\           &  9.23\pms0.17  \\
\tpipiz\nut\           &  2.54\pms0.09 & \tpi\nut\              &  9.13\pms0.15  \\
   \cline{1-2}
\pifpiz\nut            &\vspace{-2ex}& \piqpiz\nut\           &  
                                            0.03\pms0.03$^(\footnotemark[1]{^)}$ \\
\tpitpiz\nut\          &\hspace{-3mm}{$\Bigg\}$}\vspace{-2ex}\hspace{1mm}
                                         0.04 $\pm$ 0.02$^(\footnotemark[1]{^)}$  
                                       & \tpidpiz\nut\          &  0.10\pms0.02  \\
\fpipiz\nut\           &               & \fpi\nut               &  0.08\pms0.02  \\ 
   \hline
$\omega\,\pi^-$\nut$^(\footnotemark[2]{^)}$
                       &  1.83\pms0.09 & $\omega\,\pi^-$\piz\nut$^(\footnotemark[2]{^)}$
                                                                &  0.42\pms0.09  \\
$\eta\,\pi^-$\piz\nut$^(\footnotemark[3]{^)}$ 
                       &  0.17\pms0.03 & $\eta\,$\tpi\nut\      &  0.04\pms0.01  \\
      --               &      --       & $\eta\,$\pidpiz\nut\   &  0.02\pms0.01  \\
   \hline
K$^-\,$K$^0$\nut\      &  0.19\pms0.04 &    --                  &   --           \\
   \hline\hline
K$^-{\mathrm K}^+\pi^-$\nut\  
                       &  0.13\pms0.05 & K$^-{\mathrm K}^+\pi^-$\nut\  
                                                                &  0.04\pms0.04  \\
K$^0{\bar{\mathrm K}}^0\pi^-$\nut\     
                       &  0.13\pms0.05$^(\footnotemark[1]{^)}$ 
                                       & K$^0{\bar{\mathrm K}}^0\pi^-$\nut\     
                                                                &  0.04\pms0.04  \\
K${\mathrm K}^0\pi^0$\nut  
                       &  0.08\pms0.04 & K${\mathrm K}^0\pi^0$\nut\  
                                                                &  0.03\pms0.03  \\
   \hline
K$\bar{\mathrm K}\pi\pi$
                       &  0.08\pms0.08$^(\footnotemark[1]{^)}$ 
                                       & K$\bar{\mathrm K}\pi\pi$
                                        &  0.08\pms0.08$^(\footnotemark[1]{^)}$  \\
   \hline\hline 
 Total Vector          & 31.71\pms0.31 & Total Axial-Vector     & 30.42\pms0.32  \\
\hline
\end{tabular}
{\footnotesize 
\parbox{12cm}
{
\vspace{0.2cm}
$^{1}\,$The branching ratio is obtained using constraints from isospin 
        symmetry (see text and~\cite{aleph_vsf}). \\ \noindent
$^{2}\,$Through $\omega\rightarrow\pi^-\pi^+\pi^0$, 88.8\pc\ of this channel is 
        reconstructed in \tpipiz\nut. \\ \noindent
$^{3}\,$Through $\eta\rightarrow2\gamma$, 39.3\pc\ of this channel is 
        reconstructed in \pitpiz\nut. 
}} 
\end{center}
\end{table*}
\vs
-- {\bf Tau pairs} originating from Z$^0$ decays are detected utilizing 
          their characteristic collinear jet signature and the low
          multiplicity of their decays. Using the data from 1991--94,
          a total 124\,358 $\tau$ pairs is selected 
          corresponding to a detection ef\/f\/iciency of
          (78.8\pms0.1)\pc. The overall non-$\tau$ background contribution 
          in the hadronic modes amounts to (0.6\pms0.2)\pc. Details about
          the ALEPH $\tau$ pair selection are given
          in~\cite{aleph_tau,aleph_lbr,aleph_hbr}.
\vs
-- {\bf Charged particles} (electrons, muons and hadrons) are identif\/ied
          employing a maximum likelihood method to combine dif\/ferent and 
          essentially uncorrelated information measured for each individual 
          track. The procedure and the discriminating variables used in this 
          analysis are described in~\cite{aleph_lik,aleph_lbr}. The 
          calibration of charged tracks is performed using low radiating 
          \ee$\rightarrow\mu^+\mu^-$ events at beam energy and the invariant
          mass measurement of well-known, narrow resonances at low and
          intermediate energies. The resulting calibration uncertainty amounts 
          to less than 0.1\pc.
\vs
-- {\bf Photons} are reconstructed by collecting associated energetic 
          electromagnetic calorimeter (ECAL) towers, forming a {\it cluster}. 
          To distinguish genuine photons from fake photons a likelihood method 
          is applied using ECAL information, \eg, the fraction of energy 
          in the respective ECAL stacks, the transverse size of the shower 
          or the distance between the barycentre of the cluster and the closest 
          charged track. The energy calibration is performed using 
          electrons originating from Bhabha, $\tau$ and two-photon events.
          We obtain a relative calibration uncertainty of about 1.5\pc\ at
          low energy, 1\pc\ at intermediate energies and 0.5\pc\ at high
          energy.
\vs
-- The {\bf \boldmath\piz\ f\/inder} uses the virtue of a
          \piz-mass constraint f\/it to correctly attribute two reconstructed
          photons to the corresponding \piz\ decay. At higher \piz\ energy, the 
          opening angle between the boosted photons tends to become smaller 
          than the calorimeter resolution so that the two electromagnetic
          showers are often merged in one cluster. The transverse energy 
          distribution in the ECAL nevertheless allows the computation of 
          energy-weighted moments providing a measure of the two-photon 
          invariant mass. Remaining photons are considered as originating 
          from a \piz\ where the second photon has been lost.
\vs
-- The {\bf classif\/ication} of the inclusive hadronic $\tau$ decay channels 
          is performed according to~\cite{aleph_hbr} on the basis of the number 
          of reconstructed charged and neutral pions. The exclusive channels
          listed in Table~\ref{tab_va} are obtained by subtracting the 
          $\tau$ and non-$\tau$ background and the strange contribution
          from the inclusive measurements using the Monte Carlo simulation. 
          In order to extract the physical invariant mass spectra from the 
          measured ones it needs to be {\it unfolded} from the ef\/fects of 
          measurement distortion.
\vs
-- The {\bf unfolding method} used here is based on the regularised 
          inversion of the detector response matrix, obtained from
          the Monte Carlo simulation, using the {\it Singular Value
          Composition} technique. Details are published in~\cite{unfold}.

\subsection*{Systematic Errors}
\vsp
The study of systematic errors af\/fecting the measurement is subdivided
into several classes according to their origin, {\it viz.}, the photon and 
\piz\ reconstruction, the charged track measurement, the unfolding procedure 
and additional sources. Since we use an unfolding procedure based upon
a detector response matrix from the Monte Carlo simulation, the reliability
of the simulation has to be subjected to detailed 
studies~\cite{aleph_vsf,ricard,aleph_hbr}. 
\vs
In order to check the photon reconstruction in the ECAL, we studied the 
inf\/luence of calibration and resolution uncertainties as well as possible 
problems with the reference distributions of a likelihood used to veto 
fake photon candidates, of the contamination and energy distribution of 
fake photons, of the photon detection ef\/f\/iciency at threshold energies 
($E_\gamma^{\mathrm{thresh}}\,=\,300$\MeVE) and in the neighborhood of charged 
tracks.

Correspondingly, the ef\/fects of calibration and resolution uncertainties
in the measurement of charged tracks are checked, accompanied by tests of
the reconstruction ef\/f\/iciency of highly collimated multi-prong events,
and the simulation of secondary nuclear interactions.

In addition, systematic errors introduced by the unfolding procedure are
tested by comparing known, true distributions to their respective unfolded 
ones and by varying the regularisation conditions.

F\/inally, we add systematic errors due to the limited Monte Carlo
statistics and to uncertainties in the branching ratios of the respective 
$\tau$ decay modes.

\subsection*{Invariant Mass Spectra and Spectral Functions}
\vsp
The measurement procedure sketched above provides the physical invariant 
mass spectra of the measured $\tau$ decay modes including their respective
bin-to-bin covariance matrices obtained after the unfolding of the spectra
from the statistical errors and the study of systematic uncertainties. 

The exclusive vector and axial-vector $\tau$ decay channels are listed in 
Table~\ref{tab_va}. The branching ratios shown are mainly taken 
from~\cite{aleph_hbr}, including recent improvements in the precision
of the measurements of $\tau$ decays involving $\eta$ and $\omega$ 
mesons~\cite{aleph_eta,cleo_eta} as well as Kaons~\cite{kaons}.

The total vector and axial-vector current \sf~(\ref{eq_sf}) is obtained 
by summing up the exclusive \sfs\ with the addition of small contributions 
from unmeasured modes. These are taken from the {\tt KORALZ3.8, 
TAUOLA1.5}~\cite{tau_mc} $\tau$ Monte Carlo simulation (accompanied by
an inflated systematic error), as discussed below.

\subsection*{Vector Spectral Function}
\vsp
\begin{figure}[t]
  \psfile{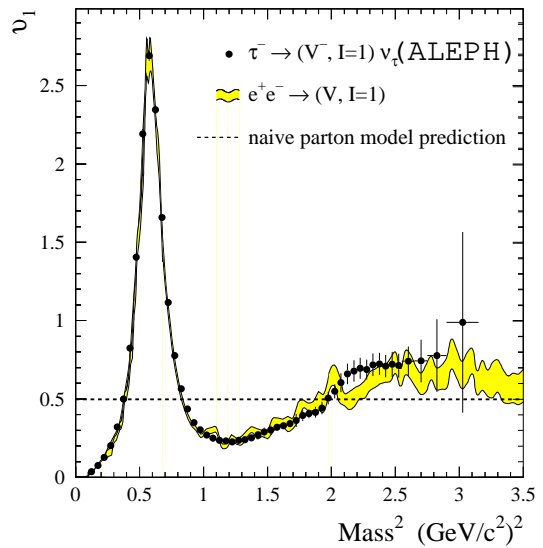}
  \caption[.]{\label{fig_vsf} 
           Total $\tau$ vector current \sf. The band shows the 
           corresponding \ee\ annihilation data obtained from
           isovector f\/inal states translated to $\tau$ decays \via\
           isospin rotation as comprehensively explained in~\cite{aleph_vsf}.
           The dashed line represents the naive parton model prediction, 
           while the QCD corrected prediction lies roughly 20\pc\ higher.}
\end{figure}
The two- and four-pion states are measured exclusively, while the six-pion
state is only partly measured. The total six-pion branching ratio has been 
determined in~\cite{aleph_vsf} using isospin symmetry. However, this 
procedure is not straight forward, as one has to consider that the 
six-pion channel is contaminated by isospin-violating \tauto$\eta$\,\tpi, 
$\eta$\,\pidpiz\nut\ decays (which were reported for the f\/irst time by 
the CLEO Collaboration~\cite{cleo_eta}).

The small fraction of the $\omega\,\pi^-$\nut\ decay channel that is not
reconstructed in the four-pion f\/inal state is added using the simulation.

Similarly, we correct for $\eta\,\pi^-$\piz\nut\ decay modes other than 
$\eta\rightarrow2\gamma$ which is classif\/ied in the \htpiz\nut\ f\/inal 
state.

The K$^-\,$K$^0$\nut\ is taken entirely from the simulation.

Using isospin constraints from $\tau$ branching ratios deduced 
in~\cite{rouge}, it can be shown~\cite{michel} that the K$\bar{\mathrm{K}}\pi$ 
channels are (78$^{+22}_{-28}$)\pc\ vector. The corresponding \sfs\ are 
obtained from the Monte Carlo simulation.

A preliminary ALEPH analysis~\cite{aleph_vsf} of $\tau$ decays into 
K$\bar{\mathrm K}\pi\pi$ f\/inal states yields a total branching ratio
of (0.17\pms0.05)\pc. However, their vector and axial-vector parts are
unknown and are therefore both estimated to be (50\pms50)\pc.
\vs
For additional details on the vector \sfs\ see Ref.~\cite{aleph_vsf,ricard}.
The total $\tau$ vector \sf\ is compared to the isospin rotated cross 
sections from isovector \ee\ f\/inal states~\cite{aleph_vsf,ricard} is 
depicted in Fig.~\ref{fig_vsf}. The agreement between both curves is 
fairly good, keeping in mind the appearance of large bin-to-bin 
correlations within the $\tau$ \sf\ as a consequence of the necessary 
regularisation during the unfolding procedure.

\subsection*{Axial-Vector Spectral Function}
\vsp
\begin{figure}[t]
  \psfile{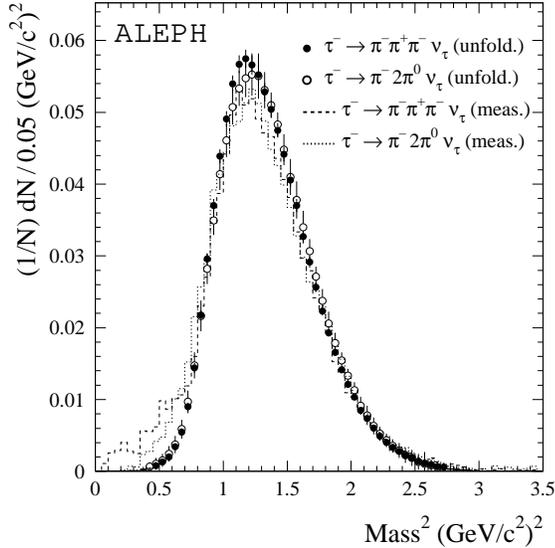}
  \caption[.]{\label{fig_a1} 
           Unfolded (physical) invariant mass-squared spectra of the $\tau$ 
           f\/inal states \tpi\nut\ and \pidpiz\nut. The dashed/dotted line
           shows the corresponding measured distribution.}
\end{figure}
\begin{figure}[t]
  \psfile{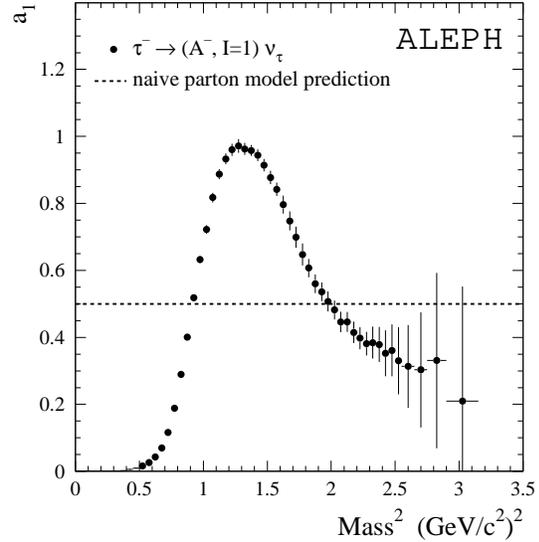}
  \caption[.]{\label{fig_asf} 
           Total $\tau$ axial-vector current \sf.
           The dashed line represents the naive parton 
           model prediction, while the QCD-corrected prediction lies 
           roughly 20\pc\ higher.}
\end{figure}
The three-pion \sfs\ are measured in both occurring f\/inal states.
Fig.~\ref{fig_a1} shows the unfolded \tpi\nut\ and \pidpiz\nut\ mass 
spectra with reasonable agreement in form and normalization. The small bump 
of the measured \tpi\nut\ spectrum at low mass (Fig.~\ref{fig_a1}) is caused 
by decays where only two charged tracks are reconstructed. Due to incomplete 
ECAL energy collection, the measured \pidpiz\nut\ distribution is slightly 
shifted to lower masses. These details are well reproduced by the Monte
Carlo simulation.  We assume in the following that both channels have 
identical spectra and use the weighted average of the distributions for 
the total axial-vector \sf. 

The f\/ive-pion \sfs\ are exclusively measured in the \tpidpiz\nut\ and
\fpi\nut\ f\/inal states. Using Pais' isospin classes~\cite{pais},
the branching fraction of \piqpiz\nut\ can be bounded entirely using the 
\fpipiz\nut\ branching fraction. It is found to be smaller than 0.055\pc. 
We take half of this upper limit with an error of 100\pc.

In analogy to the vector case, we add the small fraction of the 
$\omega\,\pi^-$\piz\nut\ decay channel that is not accumulated in the 
\tpidpiz\nut\ f\/inal state using the simulation.

Also considered are the axial-vector $\eta(3\pi)^-$\nut\ f\/inal 
states~\cite{cleo_eta} and, as in the vector case, a (50\pms50)\pc\ 
K$\bar{\mathrm K}\pi\pi$ contribution. Both \sfs\ are taken from the 
simulation accompanied by comfortable systematic errors due to the
uncertainty of the respective invariant mass distributions.
\vs
The total axial-vector \sf\ is plotted in Fig.~\ref{fig_asf}. However,
we are still not in the asymptotic region at the $\tau$ mass scale. We
may expect additional oscillations to lift the \sf\ to roughly 1.2 times 
the naive parton model prediction (that is expected asymptotically).

\subsection*{ ($v_1+a_1$) Spectral Function}
\vsp
\begin{figure}[t]
  \psfile{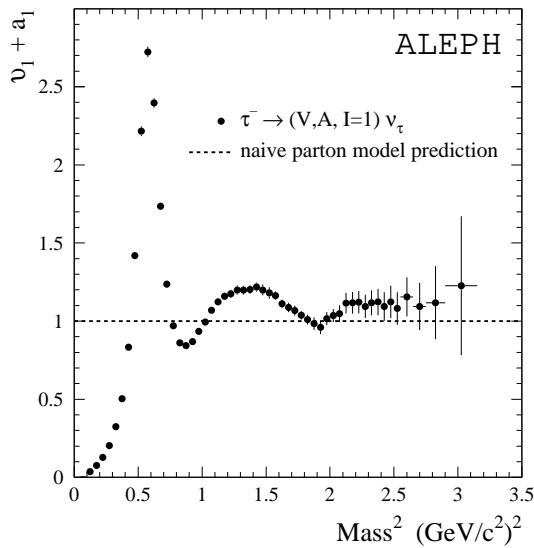}
  \caption[.]{\label{fig_vpa} 
           Inclusively measured vector plus axial-vector ($v_1+a_1$)
           \sf.  Again, the dashed line represents the naive parton model 
           prediction.}
\end{figure}
\begin{figure}[t]
  \psfile{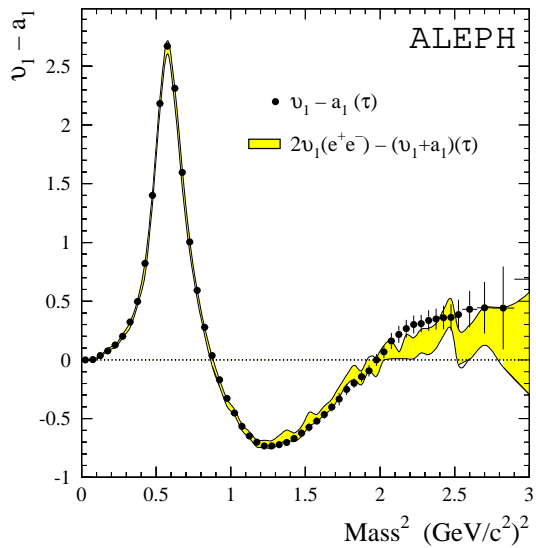}  
  \caption[.]{\label{fig_vmasf} 
           Vector minus axial-vector ($v_1-a_1$) \sf. The band shows
           the corresponding \sf\ using two times the vector \sf\ from \ee\
           annihilation data minus the inclusively measured ($v_1+a_1$) \sf\
           from $\tau$ decays.}
\end{figure}
In the favourable case of the vector plus axial-vector \sf\ we do
not have to distinguish the current properties of the respective
non-strange hadronic $\tau$ decay channels. Hence we measure the
mixture of all contributing non-strange f\/inal states as inclusively as 
possible. The dominating two- and three-pion f\/inal states are still 
measured exclusively, while the remaining contributing topologies are 
treated inclusively, \ie, without any subtraction of $\tau$-background 
originating from one of the considered decay modes. This improves the 
statistical error. Another advantage is, that we do not have to worry 
about the current properties of the K${\bar{\mathrm K}}\pi$ and 
K$\bar{\mathrm K}\pi\pi$ modes or about possible missing, \ie, unmeasured, 
$\tau$ decay modes as they are necessarily contained in the inclusive 
sample.
\vs
The ($v_1+a_1$) \sf\ is depicted in Fig.~\ref{fig_vpa}. The improvement
in precision in comparison to an exclusive sum of Fig.~\ref{fig_vsf} 
and Fig.~\ref{fig_asf} becomes obvious at higher mass-squared.
One clearly sees the oscillating \sf\ which does not seem to approach
the asymptotic limit at $s\rightarrow M_\tau^2$, which is predicted from 
perturbative QCD to lie about 20\pc\ higher than the naive parton model 
prediction.

%
%
\section{QCD CHIRAL SUM RULES}

\begin{figure*}[htb]
  \hspace{-0.1cm}
  \setlength{\epsfxsize}{8cm}
  \epsfbox{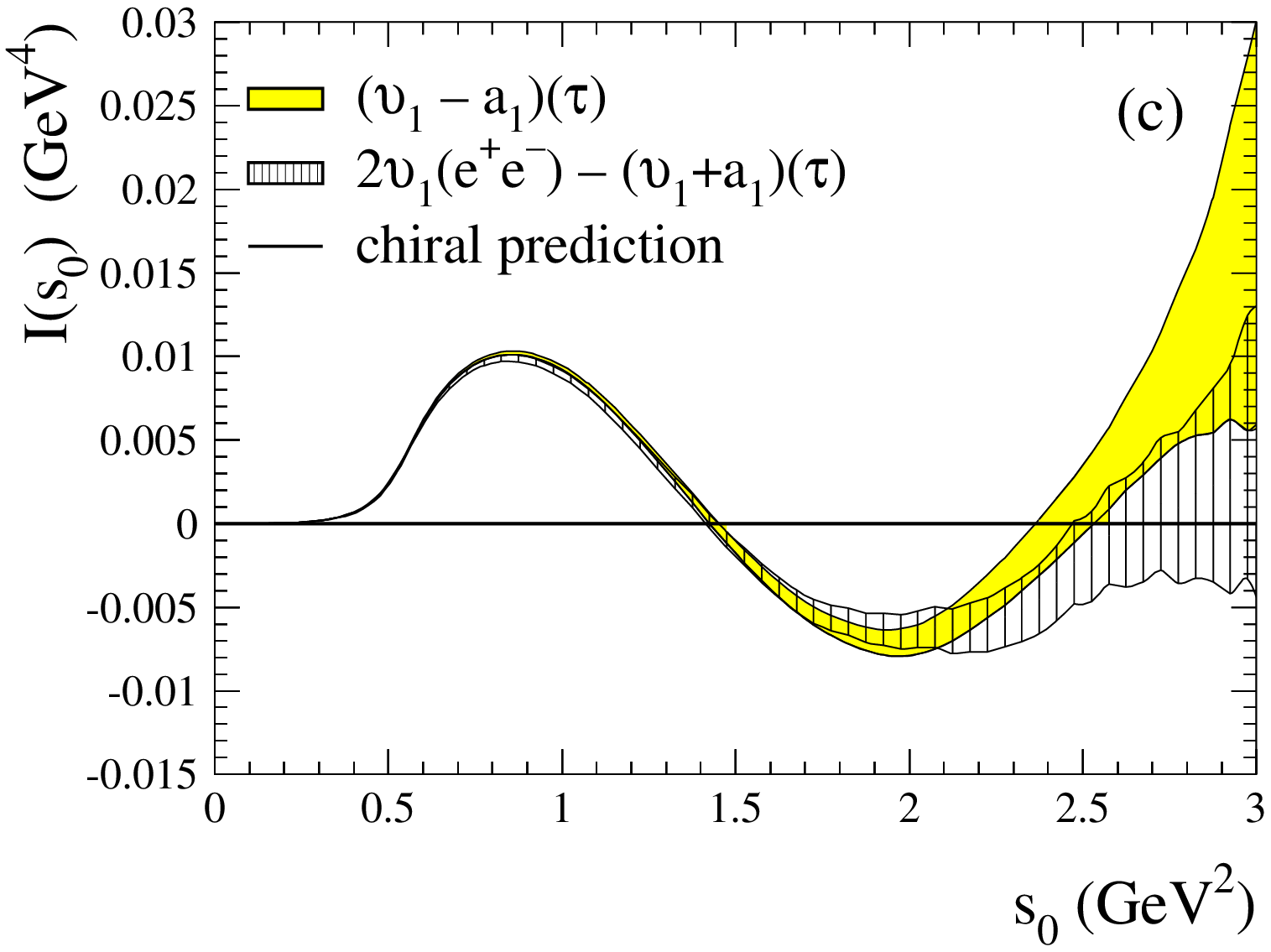}
  \setlength{\epsfxsize}{8cm}
  \epsfbox{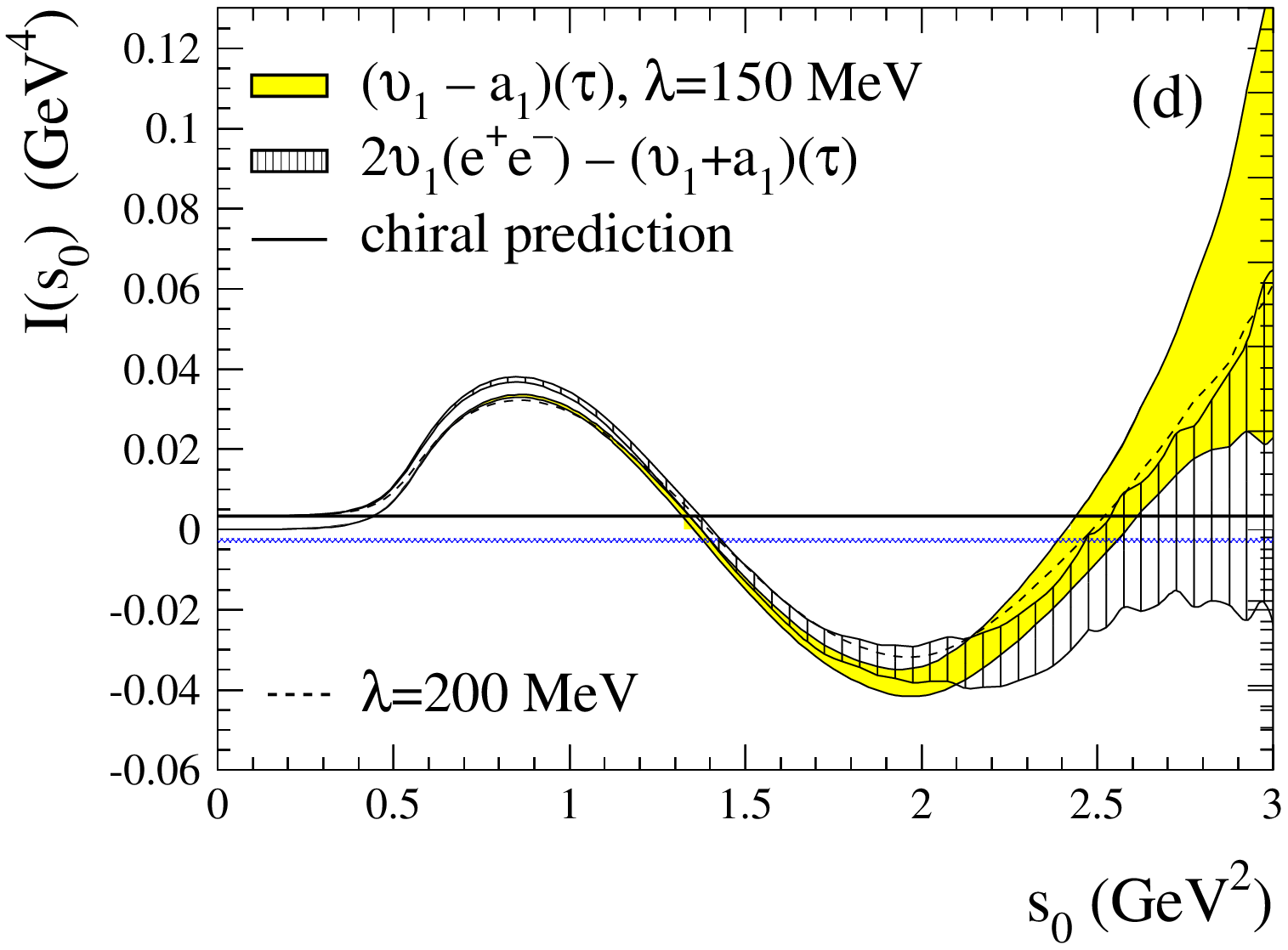}
  \vspace{-0.7cm}
  \caption[.]{\label{fig_sr}
              Sum rules corresponding to Eqs~(\ref{eq_w0})--(\ref{eq_w3}) 
              (plots: a--d) versus the upper integration bound $s_0$.}
\end{figure*}

As important application of the results obtained from the \sf\ 
analysis above, we present a study of four QCD sum rules involving
the dif\/ference of the vector and the axial-vector \sfs, each 
deduced from vector and axial-vector current conservation in the limit
of the chiral symmetry ($m_u\,=\,m_d\,=\,0$, $m_\pi\,=\,0$)\footnote
{
   We now identify the vector and axial-vector \sfs\ with the absorptive
   parts of the corresponding two-point correlation functions 
   $\Pi_{ij,V/A}^{\mu\nu}$.
}:
\beq\label{eq_w0}
   \frac{1}{4\pi^2}\hsm\intl_0^{s_0\rightarrow\infty}\hsm
                                   ds\,\frac{v_1(s)-a_1(s)}{s}
        \;=\; f_\pi^2\frac{\langle r_\pi^2\rangle}{3} - F_A~,  
\eeq
\vsm
\beq\label{eq_w1}
   \frac{1}{4\pi^2}\hsm\intl_0^{s_0\rightarrow\infty}\hsm
                                   ds\,(v_1(s)-a_1(s))
        \;=\; f_\pi^2~,  
\eeq
\vsm
\beq\label{eq_w2}
   \frac{1}{4\pi^2}\hsm\intl_0^{s_0\rightarrow\infty}\hsm 
                                   ds\,s\,(v_1(s)-a_1(s))
        \;=\;  0~,  
\eeq
\vsm
\beqn\label{eq_w3}
   \lefteqn{\frac{1}{4\pi^2}\hsm\intl_0^{s_0\rightarrow\infty}\hsm ds\,s\,
            {\mathrm{ln}}\frac{s}{\Lambda^2}\,(v_1(s)-a_1(s)) 
        \;=\;}  \nonumber \\
   & & \hspace{2cm} 
      -\frac{16\pi^2f_\pi^2}{3\alpha}\,(m_{\pi^\pm}^2-m_{\pi^0}^2)~. 
\eeqn
Eq.~(\ref{eq_w0}) is known as the Das-Mathur-Okubo (DMO) sum rule~\cite{dmo}.
It relates the given integral ($I_{\mathrm{DMO}}$) to the square of the 
pion decay constant\footnote
{
   Our def\/inition of $f_\pi$ dif\/fers from the 
   one used in~\cite{pdg} by a factor of $\sqrt{2}$.
}
$f_\pi$\,=\,(92.4\pms0.26)\MeVE~\cite{pdg} obtained from the decays
$\pi^-\rightarrow\mu^-\bar{\nu}_\mu$ and 
$\pi^-\rightarrow\mu^-\bar{\nu}_\mu\gamma$; to the pion axial-vector 
form factor $F_A$ for radiative decays 
$\pi^-\rightarrow\ell^-\bar{\nu}_\ell\gamma$; and to the pion charge
radius-squared 
$\langle r_\pi^2\rangle$\,=\,(0.431\pms0.026)~fm$^2$~\cite{finkemeier},
determined by means of a careful f\/it to space-like data~\cite{na7} 
using a two-loop chiral expansion coef\/f\/icient as additional degree 
of freedom. The error of $\langle r_\pi^2\rangle$ includes theoretical 
uncertainties. For completeness we will also express the results 
on the pion polarisability (see next paragraph) in terms of the standard 
value $\langle r_\pi^2\rangle$\,=\,(0.439\pms0.008)~fm$^2$~\cite{na7}
obtained from a one parameter f\/it to the same data~\cite{na7}. 
Eqs~(\ref{eq_w1}) and (\ref{eq_w2}) 
are the f\/irst and second Weinberg sum rules (WSR)~\cite{wsr}, where the 
pion pole in (\ref{eq_w1}) is already integrated out. When switching 
quark masses on, only the f\/irst WSR remains valid while the second 
WSR breaks down due to contributions from the dif\/ference of non-conserved 
vector and axial-vector currents of order $m_q^2/s$, leading to a 
quadratic divergence of the integral~\cite{divergence}. F\/inally, 
Eq.~(\ref{eq_w3}) represents the electromagnetic splitting of the pion 
masses~\cite{w3}.
\vs
Fig.~\ref{fig_vmasf} shows the vector minus axial-vector distribution
obtained from $\tau$ decays in addition to two times the vector \sf\ from 
\ee\ annihilation data minus the inclusive ($v_1+a_1$) measurement, yielding a 
corresponding ($v_1-a_1$) \sf\ with a slightly better precision at the end of 
the $\tau$ phase space.
The sum rules~(\ref{eq_w0})--(\ref{eq_w3}) versus the upper integration 
bound $s_0\le M_\tau^2$ are plotted in Figs.~\ref{fig_sr}a--d.
In addition to the the $\tau$ \sfs, we again show the contribution 
from \ee\ annihilation combined with the ($v_1+a_1$) inclusive measurement. 
The horizontal band depicts the corresponding chiral prediction of the 
integrals taken from~\cite{donoghue}.

We stress that only for the DMO sum rule (Fig.~\ref{fig_sr}a), where 
contributions from higher mass-squares are suppressed, does the 
saturation within the one sigma error seem to occur at the $\tau$ mass 
scale. The other sum rules (Fig.~\ref{fig_sr}b--c) are apparently not 
saturated at $M_\tau^2$ as already indicated by the non-vanishing 
($v_1-a_1$) \sf\ at the end of the $\tau$ phase space. This is due to 
a small axial-vector contribution that is clearly below the QCD 
prediction at $M_\tau^2$ as can be seen in Fig.~\ref{fig_asf}.

\subsection*{The Polarisability of the Pion}
\vsp
\begin{figure*}[htb]
  \hspace{-0.1cm}
  \setlength{\epsfxsize}{8cm}
  \epsfbox{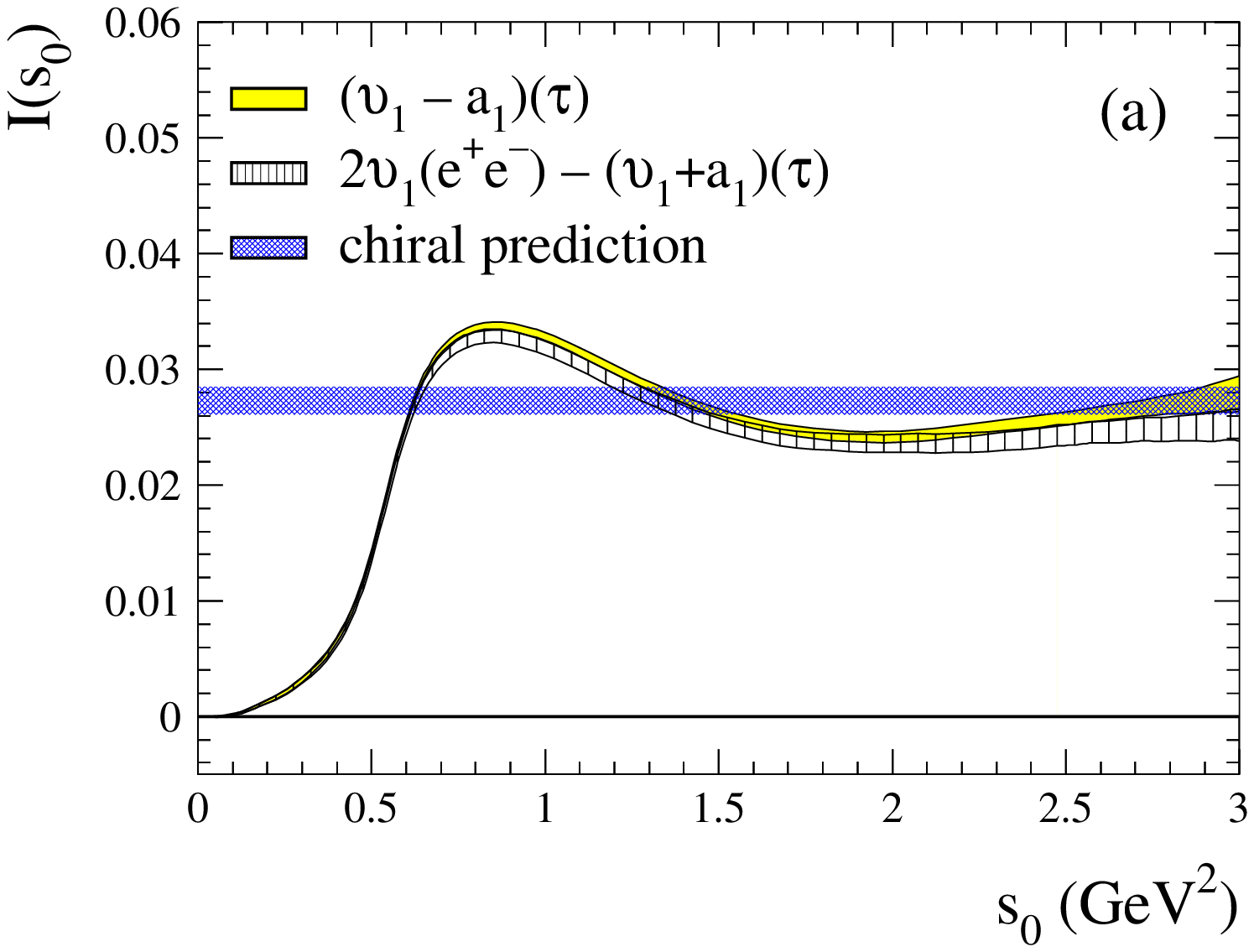}
  \setlength{\epsfxsize}{8cm}
  \epsfbox{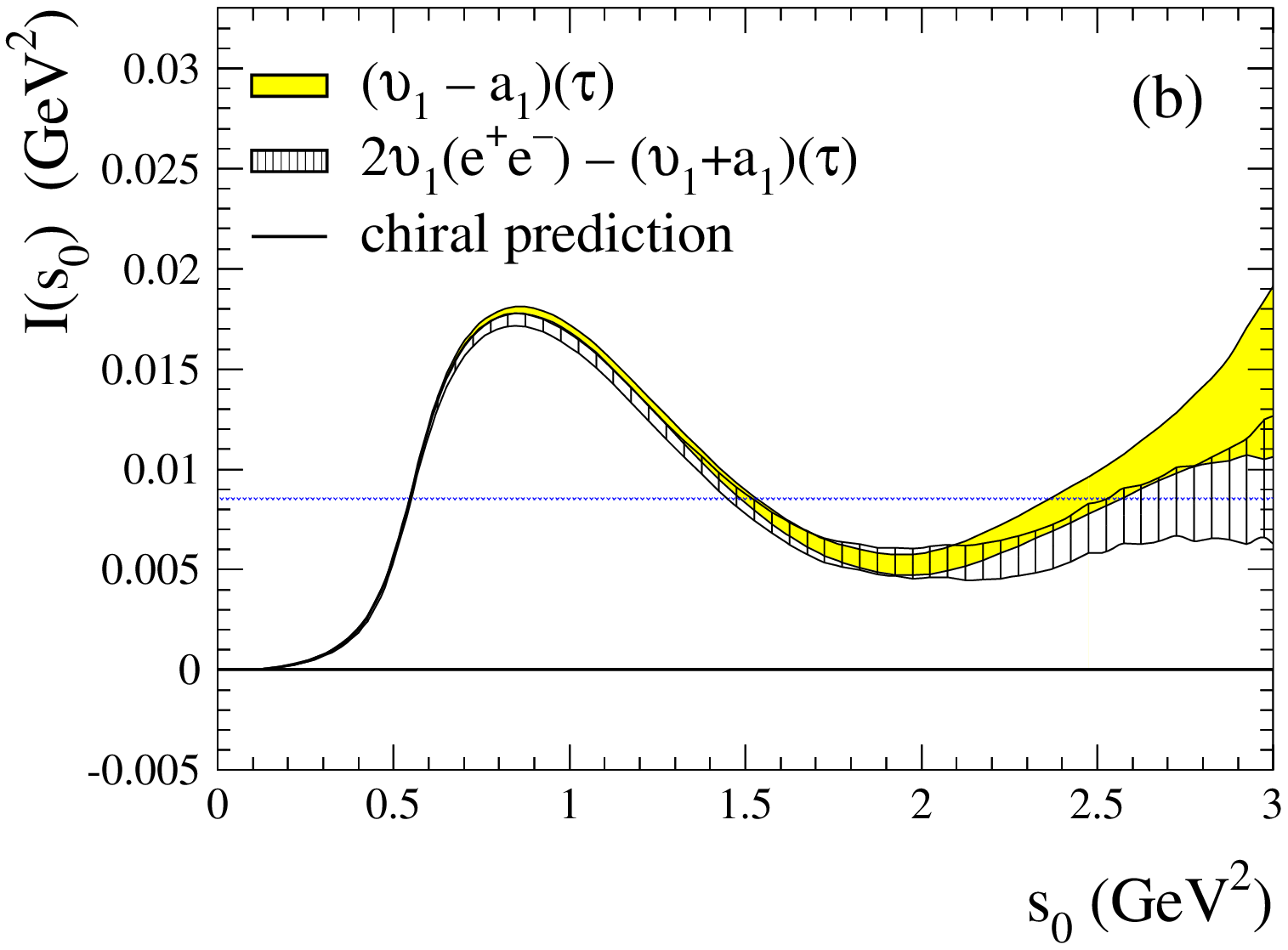}
\end{figure*}
In order to quantify the actual precision of the sum rules above, we
determine the electric polarisability\footnote
{
   The electric polarisability $\alpha_{\mathrm E}$ of a physical system 
   can be understood classically as the proportionality constant which governs
   the induction of the dipole moment $\bf p$ of a system in the presence
   of an external electric f\/ield $\bf E$: ${\bf p}=\alpha_{\mathrm E}\bf E$.
   The polarisability is an important quantity to characterize a
   particle, \ie, in probing its inner structure.
}
of the pion, given by~\cite{elpol}
\beq\label{eq_pol}
    \alpha_{\mathrm E} \;=\; \frac{\alpha F_A}{m_\pi f_\pi^2}~,
\eeq
utilizing the DMO sum rule $I_{\mathrm{DMO}}$~(\ref{eq_w0}) as proposed 
in~\cite{vato_1}. The computation of the pion axial-vector form factor $F_A$ 
is most accurate using the measurement of the ratio $\gamma\equiv F_A/F_V$
obtained from radiative pion decays $\pi^-\rightarrow e^-\bar{\nu}_e\gamma$.
Using the weighted average $\gamma$\,=\,0.46\pms0.05 of the 
measurements~\cite{gamma} as well as the CVC relation between the
pion vector form factor and the $\pi^0$ lifetime 
$|F_V|\,=\,(1/\alpha)(2\pi\tau_{\pi^0}m_{\pi^0})^{-1/2}
$\,=\,0.0132\pms0.0005~\cite{cvc_pi0,pdg}, 
we have
\beq\label{eq_polth}
    \alpha_{\mathrm E}^{\mathrm{theo}} 
        \,=\, (2.86\,\pm\,0.33)\times10^{-4}~\mathrm{fm}^3
\eeq
as a theoretical prediction.

Using the $\tau$ $(v_1-a_1)$ \sfs, we evaluate the DMO integral at the 
$\tau$ mass scale as 
$I_{\mathrm{DMO}}^\tau$\,=\,(28.0\pms1.6\pms1.1)$\times10^{-3}$, 
where the f\/irst error is the direct propagated integration error using 
the covariance matrices of the \sfs. The second error accounts for 
interpolation biases that occur because we do not integrate up
to exactly $s_0\,=\,M_\tau^2$ since the error of the \sf\ diverges at the 
very end of the $\tau$ phase space. For the $2v_1$(\ee)$-(v_1+a_1)(\tau)$ 
\sf\ we obtain 
$I_{\mathrm{DMO}}^{\tau,e^+e^-}$\,=\,(25.2\pms1.3\pms0.4)$\times10^{-3}$
with a $\chi^2\,=\,2.7/1$ when compared to the above pure $\tau$ result,
assuming both measurements to be 50\pc\ correlated. The weighted 
average (taking into account the correlation) is 
\beq\label{eq_i}
     \langle I_{\mathrm{DMO}}\rangle \,=\, 
                 (26.4 \,\pm\, 1.5)\times10^{-3}~.
\eeq
According to the prescription of Ref.~\cite{pdg}, we increased the error 
by $\sqrt{\chi^2/1}$ to account for some inconsistency. Combining (\ref{eq_w0}) 
and (\ref{eq_pol}) with the assumption that the contribution to (\ref{eq_w0}) 
for $s_0>M_\tau^2$ is negligible, \ie, the integral is saturated, 
one f\/inds that the pion polarisability is
\beq\label{eq_polexp}
     \alpha_{\mathrm E}^{\mathrm{exp}} =\, 
     (2.40\,[2.68] \,\pm\, 1.14\,[0.76])\times10^{-4}~\mathrm{fm}^3. 
\eeq
The f\/igures in brackets give the corresponding result if we use the 
standard value of 
$\langle r_\pi^2\rangle$\,=\,(0.439\pms0.008)~fm$^2$~\cite{na7} for the 
pion charge radius-squared. Both results (\ref{eq_polexp}) are in 
agreement with the chiral prediction (\ref{eq_polth}).

The authors of~\cite{vato_1} (see also~\cite{vato_2}) used the f\/irst
WSR (\ref{eq_w1}) as an additional constraint to considerably improve 
the precision and the reliability of the $I_{\mathrm{DMO}}$ evaluation 
as it naturally reduces the sensitivity to the saturation assumption.
\vs
Another approach to the solution of (\ref{eq_w0}) deals with the 
Laplace-transformed DMO sum rule, which suppresses the high energy tail
of the \sf\ in order to improve saturation at $M_\tau^2$ and to increase 
the precision of the integral~\cite{vato_1,margv}:
\beqn
   \lefteqn{\hat{I}_{\mathrm{DMO}}(M^2) \;=\; 
    \frac{1}{4\pi^2}\hsm\intl_0^{s_0\rightarrow\infty}\hsm ds\,e^{-s/M^2}\,
              \frac{v_1(s)-a_1(s)}{s}}  \nonumber \\
   & & \hspace{0.5cm}  \,+\, \frac{f_\pi^2}{M^2} 
         \,-\, \frac{C_6\langle {\cal O}(6)\rangle}{6M^6} 
         \,-\, \frac{C_8\langle {\cal O}(8)\rangle}{24M^8}
\label{eq_dmolp}
\eeqn
where $\hat{I}_{\mathrm{DMO}}(M^2)\,=\,I_{\mathrm{DMO}}$ in the limit
$M^2\rightarrow\infty$. At suf\/f\/iciently high $M^2$ the impact
of the dimension $D=6$ and dimension $D=8$ non-perturbative terms on 
$\hat{I}_{\mathrm{DMO}}(M^2)$ is small. One may use projecting
sum rules with improved saturation to determine the corresponding 
phenomenological operators:
\beqn
\label{eq_proj6}
  \lefteqn{-4\pi^2C_6\langle {\cal O}(6)\rangle \;\simeq\; 
                 \hm\intl_0^{s_0}\hm ds\,s^2(v_1(s)-a_1(s))} \nonumber\\
  & &\hspace{1.2cm} -\: \beta_1s_0\hm\intl_0^{s_0}\hm ds\,s\,(v_1(s)-a_1(s))~,
\eeqn
\vsm
\beqn
\label{eq_proj8}
  \lefteqn{4\pi^2C_8\langle {\cal O}(8)\rangle \;\simeq\; 
                 \hm\intl_0^{s_0}\hm ds\,s^3(v_1(s)-a_1(s))} \nonumber\\
  & &\hspace{1.2cm} -\: \beta_2s_0^2\hm\intl_0^{s_0}\hm ds\,s\,(v_1(s)-a_1(s))~.
\eeqn
Obviously, Eq.~(\ref{eq_proj6}) and (\ref{eq_proj8}) hold for 
$s_0\rightarrow\infty$ since the second term on the r.h.s. vanishes 
in the chiral limit by virtue of the second WSR~(\ref{eq_w2}).
Unfortunately, the gain in convergence of the sum rules obtained
from the insertion of the (even less convergent) second WSR is not 
very successful as it is accompanied by large additional errors. The
coef\/f\/icients $\beta_{1/2}$ in~(\ref{eq_proj6}) and (\ref{eq_proj8}) 
depend on the high energy tail of the ($v_1-a_1$) \sf\ which can
be expanded in powers of $s_0$. They are estimated
to be $\beta_1\approx1.5$ and $\beta_2\approx2.8$. We obtain, for 
the non-perturbative contributions, the extremely rough estimates (using 
$\tau$ data only): $C_6\langle {\cal O}(6)\rangle$=0.025\pms0.027~GeV$^6$ 
and $C_8\langle {\cal O}(8)\rangle$=$-$0.15\pms0.16~GeV$^8$. This yields 
the most precise value of $\hat{I}_{\mathrm{DMO}}$=(26.3\pms0.6) for 
$\tau$ \sfs\ at (high) $M^2$=2.6~GeV$^2$ enabling one to avoid the
large errors of the inaccurate operator terms at low energy.
\vs
A more promising solution of Eq.~(\ref{eq_dmolp}) lies in a
simultaneous f\/it of $\hat{I}_{\mathrm{DMO}}$ and the non-perturbative 
terms by means of moments on the Borel parameter $M^2$. We perform a 
$\chi^2$ f\/it using the (strongly correlated) moments
$M^2$\,=\,0.2, 0.7, 1.2, $\dots$, 3.7~GeV$^2$ in order to guarantee
suf\/f\/icient information for the constraint of the dimension 
$D=6$ and $D=8$ operators at low $M^2$ and non-biased, purely perturbative
contributions at high $M^2$ to f\/ix $\hat{I}_{\mathrm{DMO}}$ through 
the Laplace-transformed DMO integral in~(\ref{eq_dmolp}).
The f\/it converges with a $\chi^2\,=\,2.3$ over 5 degrees of freedom,
yielding (for $\tau$ data)
\beq\label{eq_i2}
     \hat{I}_{\mathrm{DMO}} \,=\,
          (25.8 \,\pm\, 0.3 \,\pm\, 0.1)\times10^{-3}~,
\eeq
which is in agreement with~(\ref{eq_i}). The second error accounts for 
estimated uncertainties induced by the saturation assumption. From the 
f\/it, the non-perturbative contributions are
$C_6\langle {\cal O}(6)\rangle$=0.0029\pms0.0002~GeV$^6$ and 
$C_8\langle {\cal O}(8)\rangle$=$-$0.0015\pms0.0003~GeV$^8$ with an
anticorrelation of nearly 100\pc.

\begin{figure}[t]
  \psfile{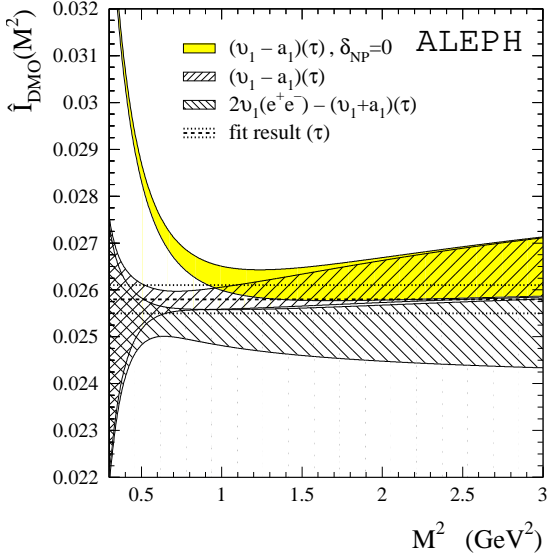}
  \caption[.]{\label{fig_borel} 
              Laplace-transformed DMO sum rule $\hat{I}_{\mathrm{DMO}}$
              as a function of the Borel parameter $M^2$ for the exclusive
              $\tau$ and the combined \ee, $\tau$ ($v_1-a_1$) \sf. The shaded
              area gives the $\tau$ integral without the non-perturbative
              contributions. It diverges hyperbolically at small $M^2$.}
\end{figure}
Fig.~\ref{fig_borel} depicts $\hat{I}_{\mathrm{DMO}}$ as a function 
of $M^2$ for the pure $\tau$ and the combined $\tau$ and \ee\ 
annihilation ($v_1-a_1$) \sfs. In this case, the pure $\tau$ \sf\ 
provides smaller errors because of the more accurate two-pion contribution 
in $\tau$ decays which is strongly weighted in~(\ref{eq_dmolp}). The 
shaded band shows the $\tau$ result without the non-perturbative terms. 
Non-perturbative contributions dominate the integral at $M^2<1$~GeV$^2$ 
and become negligible for $M^2>2.5$~GeV$^2$ as the comparison between 
the shaded and the hatched bands demonstrates. 

Using~(\ref{eq_i2}) the electric polarisability is
\beq\label{eq_polexp2}
     \alpha_{\mathrm E}^{\mathrm{exp}} =\,
     (2.68\,[2.96] \,\pm\, 0.91\,[0.32])\times10^{-4}~\mathrm{fm}^3
\eeq
with signif\/icantly lower errors compared to~(\ref{eq_polexp}) due to
the improved convergence of the Laplace-transformed DMO sum rule.
Again the f\/igures in brackets give the corresponding result when using 
the standard value of $\langle r_\pi^2\rangle$.
     
%
%
\section{THE MEASUREMENT OF \boldmath\as}

Tests of Quantum Chromodynamics and the precise measurement of the strong
coupling constant \as\ at the $\tau$ mass scale carried out for the f\/irst
time by ALEPH~\cite{aleph_as,laurent} and CLEO~\cite{cleo_as} have been
the subject of engaged discussions about theoretical and experimental
implications, accompanied by a considerable number of interesting 
publications (see, \eg,~\cite{pichledib,pert,pichtau94,aspub}). 
Following the spirit of the previous ALEPH analysis~\cite{aleph_as,laurent},
we determine \asm\ and the contributing non-perturbative terms for the 
f\/irst time using the vector, axial-vector as well as vector plus 
axial-vector \sfs\ and from a combined f\/it of all substantially 
uncorrelated input variables.

\subsection*{The $\tau$ Semileptonic Width}
\vsp
Using the analytic property of the vector/axial-vector two-point 
correlation functions $\Pi_{ij,V/A}^{(J)}$, the non-strange, semileptonic 
$\tau$ decay rates $R_{\tau,V/A}$ can be expressed as contour integrals
in the complex $s$-plane~\cite{narpich,bnp}
\beqn\label{eq_rtau1}
    \lefteqn{
      R_{\tau,V/A} \;=\; 
         6\pi i\hsm\hm\ointl_{|s|=M_\tau^2}\hsm \frac{ds}{M_\tau^2}
         \left(1-\frac{s}{M_\tau^2}\right)^{\!2}} \nonumber \\
   & & \times\left[
                   \left(1+\frac{2s}{M_\tau^2}\right)\Pi_{ud,V/A}^{(0+1)}
                   \,-\,\frac{2s}{M_\tau^2}\Pi_{ud,A}^{(0)}
             \right].
\eeqn
The advantage of this formulation is that it avoids the integration
over the whole energy-squared spectrum which includes large, uncontrolled
non-perturbative ef\/fects.

By means of the short distance OPE, 
the theoretical prediction of $R_{\tau,V/A}$ separates in the
following contributions
\beqn\label{eq_rtau2}
     \lefteqn{R_{\tau,V/A} \;=\; \frac{3}{2}|V_{ud}|^2S_{\mathrm{EW}}
                (1 \,+\, \delta_{\mathrm{EW}}^\prime \,+\,} \nonumber \\
       & & \hspace{2.6cm} \delta_{\mathrm P} \,+\, \delta_{\mathrm{mass}} \,+\,  
            \delta_{\mathrm{NP}})~,
\eeqn
where $|V_{ud}|\,=\,0.9752\,\pm\,0.0007$~\cite{pdg},  
$S_{\mathrm{EW}}\,=\,1.0194$ as in~(\ref{eq_sf}) and the small 
non-logarithmic electroweak correction 
$\delta_{\mathrm{EW}}^\prime\,\simeq\,0.0010$~\cite{delta_ew}.

The perturbative contribution $\delta_{\mathrm P}$ of the correlators 
in~(\ref{eq_rtau1}) is known to next-to-next-to-leading order $\alpha_s^3$.
Moreover, the series is resummed to all orders up to the 
unknown perturbative coef\/f\/icients of the QCD $\beta$-function.
Hence the convergence of the series is reinforced and the dependence
on the scale and the choice of the renormalization scheme is 
reduced~\cite{pert}. 

The leading quark mass correction $\delta_{\mathrm{mass}}$ 
in~(\ref{eq_rtau2}) is safely neglected for $u$ and $d$ quarks 
($\approx\,-0.08$\pc~\cite{pichtau94}).

For the non-perturbative contribution $\delta_{\mathrm{NP}}$ we use
the SVZ-approach~\cite{svz} in terms of a power series 
of $M_\tau$. Long-distance QCD ef\/fects are thereby separated into
unpredictable, non-perturbative contributions, absorbed in dimensional, 
phenomenological operators ${\cal O}(D)$ and into short distance parts 
$C_D$ (Wilson Coef\/f\/icients), calculable within perturbative QCD:
\beq\label{eq_ope}
    \delta_{\mathrm{NP}} \;= \hm\sum_{D=4,6,8}\hsm\delta^{(D)} \;=
       \hm\sum_{D=4,6,8}\hm\hm\hm C_D\frac{{\cal O}(D)}{(-M_\tau^2)^{D/2}}~.
\eeq
The f\/irst operator that involves non-perturbative physics appears at 
$D=4$ and is linked to the gluon condensate.

\subsection*{Spectral Moments}
\vsp
To detach the measurement from theoretical constraints on the operators
${\cal O}(D)$, it is convenient to f\/it simultaneously \asm, 
${\cal O}(4)$, ${\cal O}(6)$ and ${\cal O}(8)$. In order to incorporate
new experimental information, we exploit the explicit shape of the 
(normalized) vector/axial-vector invariant mass spectrum 
$(1/N_{V/A})\,dN_{V/A}/ds$ by means of the spectral moments
\beqn
    \lefteqn{D_{V/A}^{kl} \;=\; \hm\intl_0^{M_\tau^2}\hm ds\,
               \left(1-\frac{s}{M_\tau^2}\right)^{\!k}
               \left(\frac{s}{M_\tau^2}\right)^{\!l}} \nonumber \\
     & &    \hspace{3cm}\times\frac{1}{N_{V/A}}\frac{dN_{V/A}}{ds}~,
\label{eq_moments}
\eeqn
where the choice of $k=1$ and $l=0,1,2,3$ provides four additional 
(correlated) degrees of freedom for each vector and axial-vector f\/it. 

\subsection{Theoretical Uncertainties}
\label{sec_theo_unc}
\vsp
The estimate of theoretical uncertainties which is included in the 
f\/it takes into account uncertainties in the physical 
constants used and tiny contributions from quark mass corrections. In 
addition, the ef\/fect of unknown higher order perturbative contributions
is estimated by varying the leading unknown coef\/f\/icient $K_4$
between zero and $K_4\approx2|K_3(K_3/K_2)|\approx50$ (to be compared, 
\eg, with the experimental estimate $K_4$\,=\,27\pms5~\cite{dibtau94}). The 
ef\/fect of the unknown $\beta_4$ coef\/f\/icient of the $\beta$-function 
is estimated in a similar way but turns out to be small. The uncertainty
from the ambiguity of the renormalization scale $\mu$ is rather small;
nevertheless it is considered here. We additionally include rough 
estimates of higher order non-perturbative operators which are not 
measured here that lead to small contributions to the theoretical error.

The dominant theoretical uncertainty for the prediction of $R_{\tau,V/A}$ 
comes from the unknown $K_4$ coef\/f\/icient.

\subsection*{Results}
\vsp
Computing the sum of the branching fractions from Table~\ref{tab_va} of 
the exclusive contributions to the vector and axial-vector \sf\ yields 
the semileptonic widths
\beqn
\label{eq_rtauv}
    R_{\tau,V}      &=& 1.782 \,\pm\, 0.018 \;\,(\pm\, 0.016)~, \\
\label{eq_rtaua}
    R_{\tau,A}      &=& 1.711 \,\pm\, 0.019 \;\,(\pm\, 0.016)~, \\
\label{eq_rtauvpa}
    R_{\tau,V+A+S}  &=& 3.649 \,\pm\, 0.013 \;\,(\pm\, 0.036)~,
\eeqn
where the f\/irst errors are from experimental origin while the second
ones (in parentheses) represent uncertainties in the theoretical 
predictions as described in Section~\ref{sec_theo_unc}. 
The value of~(\ref{eq_rtauvpa}) is obtained via universality from
the $\tau$ branching ratios into $e^-\bar{\nu}_e$\nut\ and
$\mu^-\bar{\nu}_\mu$\nut\ and the $\tau$ lifetime which are all taken 
from~\cite{br_e,pdg}. The relative rise of the theoretical error 
in~(\ref{eq_rtauvpa}) compared to~(\ref{eq_rtauv}) and (\ref{eq_rtaua}) 
comes from the additional uncertainties in the strange sector. It 
can be reduced to $\Delta^{\mathrm{th}}R_{\tau,V+A+S}\,=\,0.034$ 
when f\/itting simultaneously the mass of the strange quark via 
$R_{\tau,S}$~\cite{michel}. 
\vs
\begin{table}[th]
  \begin{center}
\parbox{\columnwidth}
{
  \caption[.]{\label{tab_moments}
              Spectral Moments of vector ($V$), axial-vector ($A$) and
              vector plus axial-vector ($V+A$) $\tau$ decay modes.
              The f\/irst error lines give the total experimental error
              including statistical and systematic uncertainties while
              the second error lines account for uncertainties in the 
              theoretical prediction.}
\vspace{0.3cm}
}
{\small
  \begin{tabular}{|c||c|c|c|c|} \hline 
 ALEPH          &   $l=0$   &   $l=1$   &   $l=2$   &   $l=3$   \\ \hline\hline
 $D_V^{1l}$     &   0.7159  &   0.1689  &   0.0532  &   0.0227  \\ \hline
 $\Delta^{\mathrm{exp}} D_V^{1l}$      
                &   0.0034  &   0.0006  &   0.0007  &   0.0006  \\
 $\Delta^{\mathrm{th}} D_V^{1l}$     
                &   0.0037  &   0.0035  &   0.0004  &   0.0002  \\ \hline\hline
 $D_A^{1l}$     &   0.7205  &   0.1471  &   0.0639  &   0.0303  \\ \hline
 $\Delta^{\mathrm{exp}} D_A^{1l}$      
                &   0.0033  &   0.0009  &   0.0005  &   0.0004  \\
 $\Delta^{\mathrm{th}} D_A^{1l}$     
                &   0.0037  &   0.0031  &   0.0005  &   0.0003  \\ \hline\hline
 $D_{V+A}^{1l}$ &   0.7177  &   0.1581  &   0.0585  &   0.0265  \\ \hline
 $\Delta^{\mathrm{exp}} D_{V+A}^{1l}$  
                &   0.0022  &   0.0006  &   0.0004  &   0.0004  \\
 $\Delta^{\mathrm{th}} D_{V+A}^{1l}$ 
                &   0.0037  &   0.0033  &   0.0004  &   0.0002  \\ \hline
  \end{tabular}
}
  \end{center}
\end{table}
\begin{table*}[th]
  \begin{center}
  \parbox{13cm}
  {
  \caption[.]{\label{tab_results}
              F\/it results of \asm\ and the OPE non-perturbative 
              contributions~(\ref{eq_ope}) from various input parameters: 
              the $V$ ($A$) f\/it uses $R_{\tau,V}$ ($R_{\tau,A}$) and the 
              corresponding moments. The ($V+A$) moments are f\/itted
              together with $R_{\tau,V+A+S}$. The combined f\/it exploits 
              the information of $R_{\tau,V}$, $R_{\tau,A}$, $R_{\tau,V+A+S}$
              and $R_{\tau,S}$ as well as the vector and axial-vector moments.}
  \vspace{0.3cm}
  }
  {\small
  \begin{tabular}{|c||c|c|c||cc|} \hline 
     & & & & \mc{2}{c|}{Combined F\/it}                  \\
  \rs{ALEPH}     &\rs{Vector ($V$)} &\rs{Axial-Vector ($A$)}&  \rs{$V\,+\,A$}    
                 & Vector           & Axial-Vector       \\ \hline \hline
  \asm\          &  0.354\pms0.020  &  0.358\pms0.023   &  0.348\pms0.017  
                 & \mc{2}{c|}{0.351\pms0.016}            \\ \hline
  $\delta^{(4)}$ & (0.1\pms0.4)\TT  &($-$1.1\pms0.7)\TT & $-$(0.8\pms1.2)\TT
                 & \mc{2}{c|}{($-$0.7\pms0.6)\TT}        \\
  $\delta^{(6)}$ &  0.029\pms0.004  & $-$0.028\pms0.004 & 0.002\pms0.005  
                 &  0.023\pms0.003  & $-$0.024\pms0.003  \\ 
  $\delta^{(8)}$ &($-$9.0\pms1.1)\TT& (7.7\pms1.0)\TT   & ($-$1.0\pms1.0)\TT 
                 &($-$8.8\pms1.0)\TT& (8.3\pms1.0)\TT    \\ \hline
  $\chi^2/$d.o.f.& 0.4/1            & 0.4/1             & 0.1/1           
                 & \mc{2}{c|}{8.0/5}                     \\ \hline
  \end{tabular}
  }
  \end{center}
\end{table*}
Table~\ref{tab_moments} lists the measured values of the spectral
moments~(\ref{eq_moments}) together with the experimental errors and 
the uncertainties from the theoretical predictions for the vector, axial-vector 
and vector plus axial-vector \sfs. Note that all the moments obtained from 
the same \sf\ are strongly correlated. The non-perturbative 
terms~(\ref{eq_ope}) are adjusted in the f\/it so that they do not 
contribute to the theoretical errors. In particular, theoretical 
uncertainties dominate the total error for the lower moments ($l=0,1$) 
due to the important error contribution from the missing $K_4$ 
coef\/f\/icient of the perturbative expansion. 

\begin{figure}[t]
  \psfile{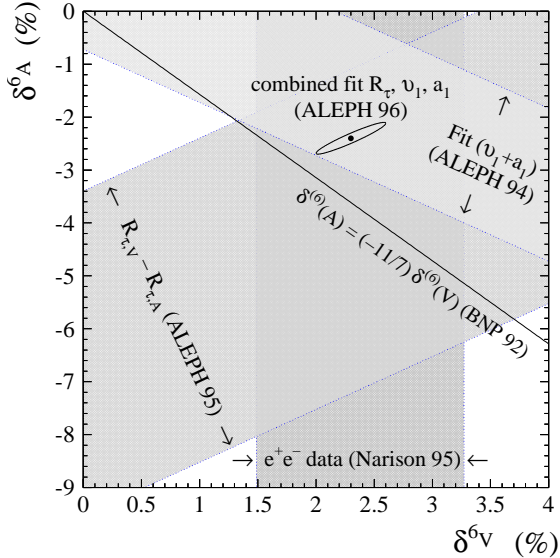}
  \caption[.]{\label{fig_delta6} 
              Non-perturbative contribution $\delta_6$ from 
              Eq.~(\ref{eq_ope}) to the $\tau$ hadronic width of
              vector and axial-vector f\/inal states. The ellipse 
              depicts the ALEPH result obtained in the combined f\/it.}
\end{figure}

\begin{figure}[t]
  \psfile{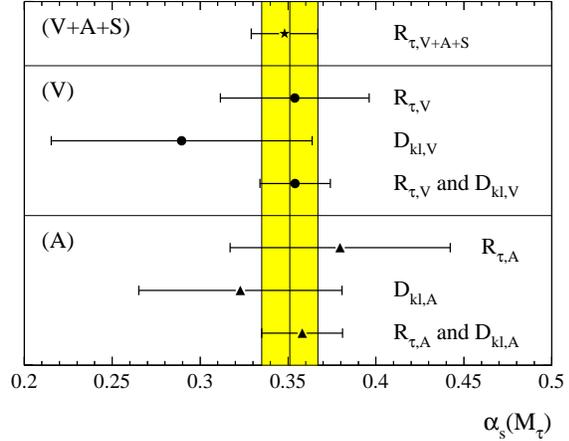}
  \caption[.]{\label{fig_alpha_s} 
              Results for \asm\ using $R_\tau$ only, the moments only
              and the combined information from vector and axial-vector
              $\tau$ decays. Additionally, the result from 
              $R_{\tau,V+A+S}$ only is plotted. The shaded band depicts
              \asm\ from the combined vector and axial-vector f\/it as
              given explicitly in Table~\ref{tab_results}.}
\end{figure}

The results of the various f\/its of \asm\ and the non-perturbative terms
of dimension $D=4,6,8$ are given in Table~\ref{tab_results}.
The combined f\/it uses as input parameters $R_{\tau,V}$, $R_{\tau,A}$, 
$R_{\tau,V+A+S}$, $R_{\tau,S}$ and the vector and axial-vector moments. 
As mentioned above, the additional parameter 
$R_{\tau,S}$\,=\,0.1560\pms0.0064~\cite{michel} is used to adjust the 
strange quark mass in order to reduce the theoretical uncertainty of the 
$R_{\tau,V+A+S}$ prediction. The gain from the separation of vector and 
axial-vector channels compared to the inclusive $V+A$ f\/it becomes 
obvious in the adjustment of the leading non-perturbative contributions
of dimension $D=6$ and $D=8$, which cancel in the inclusive sum. The 
information for their accurate determination comes mainly from the high
$l=3,4$ moments. Fig.~\ref{fig_delta6} presents the total $D=6$ contribution
$\delta^{(6)}\,=\,C_6\langle {\cal O}(6)\rangle/M^6$ to the vector and 
axial-vector hadronic widths as obtained from the combined f\/it. It is 
shown in comparison to recent experimental and theoretical constraints 
taken from~\cite{aleph_as,aleph_hbr,bnp,narison}.
\vs
Fig.~\ref{fig_alpha_s} depicts the constraints
of $R_\tau$ and the spectral moments on \asm\ when using standard values
for the non-perturbative contributions~\cite{bnp} with the addition of
large systematic uncertainties. They are in excellent agreement with 
the best \asm\ measurement obtained from the combined f\/it. The fact that
the normalization of the respective spectra ($R_\tau$) and their pure 
shapes (spectral moments) independently give consistent results is a 
remarkable conf\/irmation of the reliability of the \as\ measurement 
using $\tau$ decays and the underlying SVZ approach~(\ref{eq_ope}).
\vs
The evolution~\cite{wetzel,bern} of the best \asm, taken from the combined 
f\/it, to the Z boson mass yields the f\/inal result
\beqns
   \alpha_s(M_{\mathrm Z}) \;=\; 
        0.1219\,\pm\,0.0006\,\pm\,0.0015\,\pm\,0.0010~.
\eeqns
The f\/irst error accounts for the experimental uncertainty, the 
second one gives the uncertainty of the theoretical prediction of
$R_\tau$ and the spectral moments, while the last error stands for
possible ambiguities in the evolution due to uncertainties of the 
matching scales of the quark thresholds (0.0010) and ef\/fects
associated with the truncation of the RGE at next-to-next-to-leading
order (0.0003).

\section{CONCLUSIONS}

In this talk, we presented new results on $\tau$ hadronic \sfs, measured
by the ALEPH Collaboration at LEP, and their implications for QCD.
With small inputs from unmeasured $\tau$ decay modes, the vector
and axial-vector contributions could entirely be separated providing
accurate experimental access to chiral sum rules and non-perturbative
phenomenological QCD. We used moments of the Laplace-transformed
Das-Mathur-Okubo (DMO) sum rule for a precise determination of the
DMO integral and, through it, of the electric polarisability of the 
pion. F\/inally, we used both vector and axial-vector \sfs\ for
an exhaustive study of \as\ and leading non-perturbative terms 
at the $\tau$ mass scale. For the f\/irst time, the latter could be 
f\/ixed with surprising precision from $\tau$ decays. A combined 
f\/it of all output variables yielded 
\as$(M_{\mathrm Z})$\,=\,0.1219\pms0.0019 after the evolution from
the $\tau$ to the Z mass.

\section*{ACKNOWLEDGEMENTS}

I gratefully acknowledge the organizing committee and especially J.~Smith
for their ef\/forts to realize this interesting conference at such a
splendid place in the Rocky Mountains. I would like to thank Michel Davier
and Ricard Alemany for the fruitful collaboration during the last years of
our work on $\tau$ \sfs.

\end{document}